# Interband resonant high-harmonic generation by valley polarized electron-hole pairs

Naotaka Yoshikawa[1,2,*], Kohei Nagai[1], Kento Uchida[1], Yuhei Takaguchi[3], Shogo Sasaki[3], Yasumitsu Miyata[3], and Koichiro Tanaka[1,2,†]

[1] *Department of Physics, Graduate School of Science, Kyoto University, Sakyo-ku, Kyoto 606-8502, Japan*

[2] *Institute for Integrated Cell-Material Sciences (WPI-iCeMS), Kyoto University, Sakyo-ku, Kyoto 606-8501, Japan*

[3] *Department of Physics, Tokyo Metropolitan University, Hachioji, 192-0397, Japan.*

[*] yoshikawa@thz.phys.s.u-tokyo.ac.jp

[†] kochan@scphys.kyoto-u.ac.jp

**ABSTRACT**

We demonstrated nonperturbative high harmonics induced by intense mid-infrared light up to 18th order that well exceed the material bandgap in monolayer transition metal dichalcogenides. The intensities of the even-order high-harmonic radiation did not monotonically decrease as the harmonic order increased. By comparing the high harmonic spectra with the optical absorption spectra, we found that the enhancement in the even-order high harmonics could be attributed to the resonance to the band nesting energy. The symmetry analysis shows that the valley polarization and anisotropic band structure lead to polarization of the high-harmonic radiation under excitation with the polarization along the zigzag direction. We also examined the possible recombination pathways of electrons and holes by calculating their dynamics in real and momentum spaces based on three-step model in solids. It revealed that, by considering the electrons and holes generated at neighboring lattice sites, the electron-hole polarization driven to the band nesting region should contribute to the high harmonic radiation. Our findings open the way for attosecond science with monolayer materials having widely tunable electronic structures.

**TEXT**

Research on generation of high harmonics (HHG) from atomic or molecular gases has produced coherent broadband radiation in the extreme ultraviolet region and attosecond pulses[1,2]. Moreover, HHG has recently been reported from several solid-state materials[3-12]. Nonperturbative HHG in solids can be achieved with a driving laser with a peak intensity as low as $10^{12}$ W/cm$^2$, which is much smaller than that required in the gas phase, suggesting the potential for a compact, highly efficient attosecond light source. In addition, solid-state HHG can capture the electronic properties of materials as tomographic imaging of molecular gases has been demonstrated by high-harmonic spectroscopy[13-15]. High harmonic spectroscopy in solids shows the possibility of probing the electronic band structure as it relates to crystal symmetry and interatomic bonding[3,4,7,10,11]. A unified understanding of the mechanism of HHG in solids is necessary for the application of high harmonic spectroscopy. Although several theoretical models have been proposed for solid-state HHG in terms of interband polarization and intraband electron dynamics as well as their interplay[3-11,16-19], the underlying mechanism is still under debate. It will be important to gain an understanding of HHG in single-atomic-layer solids, because propagation effects such as the phase matching condition, which would otherwise obscure intrinsic features of HHG in bulk crystals, do not affect ideal two-dimensional systems.

HHG from monolayer graphene was recently reported[20-22], and its observed anomalous ellipticity dependence was able to be reproduced by a fully quantum mechanical theory in which both intraband and interband contributions are included[18]. HHG was also observed from monolayer MoS$_2$, which is one of the monolayer transition metal dichalcogenides (TMDs)[23]. Monolayer TMDs, which have a finite bandgap in contrast to graphene, have attracted much attention for their exotic properties such as enhancement of luminescence derived from their indirect-to-direct bandgap transition from bulk to monolayer form[24,25], extremely large exciton binding energy[26], and valley pseudospin physics arising from inversion symmetry breaking and a large spin-orbit interaction[27,28]. Liu et al. found that the even-order high harmonics are predominantly polarized perpendicular to the pump laser and attributed this to the intraband anomalous transverse current arising from the material's Berry curvature[23]. On the other hand, a recent experiment with an optical pump showed that the interband contribution dominates the intraband one in bulk $ZnO$[29]. Liu et al. also showed that the even orders of HHG are enhanced at much higher energies compared with the lowest exciton or bandgap energy, while the odd orders are monotonically suppressed at higher energies. The enhancement at higher orders and its relation to the perpendicular polarization of even-order HHG is not yet understood.

In this study, we systematically investigated HHG from four kinds of monolayer TMDs with different bandgap energies ($MoSe_2$, $WSe_2$, $MoS_2$, and $WS_2$) to examine the resonance effect and polarization of HHG in monolayer TMDs. By comparing the HHG and optical absorption spectra of the monolayer TMDs, we found that HHG is enhanced when it is in resonance with the optical transition due to band nesting, indicating that the interband polarization mainly contributes to the even-order HHG. A simple calculation that is an extension of the three-step model for electron-hole pairs in solids shows that the electron-hole pairs driven to the band nesting region significantly contributes to the resonance enhancement, and anisotropic driving of electron-hole polarizations around the $K$ and $K'$ points determines polarization selection rules of HHG.

## Results

**High harmonic spectra from monolayer TMDs**

We excited the monolayer TMDs by using mid-infrared pulses (0.26 eV photon energy, 1.7 TW/cm$^2$ peak intensity) and detected HHG in the near-infrared to ultraviolet energy region (see the Methods and Supplementary Figure S1). Figure 1 shows the high harmonic spectra from fifth to sixteenth order of (a) MoSe$_2$, (b) WSe$_2$, (c) MoS$_2$, and (d) WS$_2$ monolayers (see the Methods and Supplementary Figure S2 for the full spectrum up to 18th harmonics). We set the excitation polarization of the mid-infrared pulses in parallel with the zigzag direction (M-M (M=Mo, W) direction) (see Supplementary Note 1). The polarization of the even-order harmonic radiation was perpendicular to that of the incident laser light, while the polarization of the odd-order harmonics was parallel.[23] In contrast, when we set the excitation polarization of the mid-infrared pulses in parallel with the armchair direction, both of odd-order and even-order harmonics have parallel polarization to the excitation polarization. Detailed results are shown in Supplementary Figure S3.

**Interband resonant enhancement of HHG**

Figures 2(a)-2(d) show the polarization-unresolved emission intensities of the even-order harmonics, and Figs. 2(e)-2(f) the optical absorption spectra. In the absorption spectra, the peaks labeled A and B are attributed to so-called A and B excitons, consisting of electrons and holes localized in the *K* valleys in momentum space. In addition, the band structures of the monolayer TMDs have a band nesting region, causing van Hove singularities in the joint density of states[30,31]. The band nesting results in strong optical absorption at higher photon energies (labeled C and D in the figure)[32,33]. As shown in Fig. 2(a), the intensity of the 10th harmonic in monolayer MoSe$_2$ is larger than that of the 8th, and the intensity of the 14th harmonic is larger than that of the 12th.

Interestingly, the energies of the 10th and 14th harmonics coincide with the absorption peaks C and D. This is not only the case in MoSe$_2$. Figures 2(b)-2(d) also show enhancements at the band nesting energies for WSe$_2$, MoS$_2$, and WS$_2$. The observed enhancement in resonance with the interband optical transition strongly suggests that the interband contribution is dominant in even-order HHG. On the other hand, the intensities of the odd orders monotonically decrease with increasing order as shown in Figs. 2(i)-2(l). However, if looking carefully to Figs. 2(i)-2(l), the shoulder-like structures are clear around the C and D transitions. This suggests that the enhancement should exist even in the odd-order harmonics but is hidden by the contribution from the intraband process. The intraband process may show no resonant enhancement around the band nesting region. The experimental data imply that the intraband contribution is much larger than the interband one for the odd-order harmonics in the low energy region, but the interband polarization may substantially contributes it at the resonance energies.

**Polarization selection rules of HHG**

The resonant enhancement of HHG indicates that the interband polarization mainly contributes to even-order HHG. The question is how the perpendicular polarization arises in the even-order harmonics. The transverse current due to a material's Berry curvature might not contribute to interband HHG. Here, we will discuss the mechanism of HHG in monolayer TMDs on the basis of a semi-classical three-step model in solids. The discussion is fundamentally the same as the one on the polarization of high-order sideband generation in Ref. [34]. In our experiment, the incident light was linearly polarized; i.e., it was a linear combination of left- and right-hand circular polarized light. In the strong-field regime, therefore, electron-hole pairs are generated coherently by Zener tunneling at the $K$ and $K'$ points in the two-dimensional hexagonal Brillouin zone. We will assume that electron-hole pairs are created when the electric field of the incident light reaches positive or negative maxima. Since the generated electrons and holes are accelerated differently in the conduction and valence bands, their motions are asymmetric in real space (Fig. 3c). When the electron and hole meet again in real space (the circle in Fig. 3d), electron-hole recombination occurs and HHG is emitted. The electron-hole pairs created at the positive peak field ($t = 0$ in Fig. 3d) under excitation with a polarization along the zigzag direction (Fig. 3a) have anisotropic driving processes: those created at the $K$ point are driven in the $K - \Gamma - K'$ direction, while those created at the $K'$ point are driven in the $K' - M - K$ direction. This anisotropy in turn causes anisotropic acceleration and recombination dynamics in the $K$ and $K'$ bands ($0 < t < T/2$, where $T$ is the period of the incident light). The dynamics of the electron-hole pairs generated at the negative peak field ($t = T/2$) are the reverse of those generated at the positive peak field ($T/2 < t < T$). On the other hand, when the excitation light is polarized along the armchair direction, the electron-hole pairs generated at the positive peak field and those generated at the negative peak field have the same dynamics in the $K$ and $K'$ bands (Fig. 3e).

Next, we show how the anisotropic and alternating nature of the electron-hole dynamics gives rise to a polarization selection rule for HHG. The electric field of HHG is the sum of the left- and right-hand circular polarized light:

$$\boldsymbol{E}^{\text{HHG}}(t) = E_+^{\text{HHG}}(t)\boldsymbol{\sigma}^+ + E_-^{\text{HHG}}(t)\boldsymbol{\sigma}^- \qquad (1)$$

where $E_\pm^{\text{HHG}} = (E_x^{\text{HHG}} \pm iE_y^{\text{HHG}})/\sqrt{2}$ are the amplitudes of the $\boldsymbol{\sigma}^\pm$ components. Here, we regard the $x$ axis as the zigzag direction and suppose mid-infrared excitation with the polarization of the zigzag direction: $\boldsymbol{E}_{zigzag}^{MID}(t) = E_x(t)(\boldsymbol{\sigma}^+ + \boldsymbol{\sigma}^-)/\sqrt{2}$. We only consider the ballistic driving process for HHG since the electrons and holes losing their initial coherence do not contribute to high harmonics. We assume that the relative phase between electron-hole pairs generated at $K$ and $K'$ points is conserved in the whole process[19]. Even if their spin states change by driving process, the following discussion is valid when the initially generated, relative phase is not lost. The mirror symmetry of the system and the periodicity of the incident light with the polarization of the zigzag direction follows the relation (see Supplemental Note 4 for the detailed discussion):

$$E^{\mathrm{HHG}}_{\pm}(t) = -E^{\mathrm{HHG}}_{\mp}(t + \frac{T}{2}) \qquad (2)$$

This directly gives the selection rules:

$$E^{\mathrm{HHG}}_{x}(t) = -E^{\mathrm{HHG}}_{x}(t + \frac{T}{2}) \qquad (3)$$

and

$$E^{\mathrm{HHG}}_{y}(t) = E^{\mathrm{HHG}}_{y}(t + \frac{T}{2}) \qquad (4)$$

By considering $\boldsymbol{E}^{\mathrm{HHG}}(t) = \sum_n \boldsymbol{E}_n e^{in\omega t} = \sum_n [E_x{}^n \hat{\boldsymbol{x}} + E_y{}^n \hat{\boldsymbol{y}}] e^{-in\omega t}$, equations (3) and (4) tell us that the odd-order HHG has parallel polarization and even-order HHG have perpendicular polarizations to the incident light. The polarization selection rules of the HHG obtained here explain the results of our experiment and Ref. [23]. We also obtain the polarization selection rule of HHG with the armchair excitation in the similar way (See Supplemental Note 4).

**Calculation of dynamics of electrons and holes by semi-classical treatment**

To understand the resonant effect of interband HHG in the band nesting energy region under the zigzag excitation, it is worthwhile to know where the recombined electron-hole pairs are in momentum space. Here, we calculated the electron-hole dynamics in real and momentum space for monolayer MoS$_2$ on the basis of the three-step model. The valence and conduction band were given by a tight-binding model without spin-orbit couplings[35,36] (see the Supplementary Note 3). We supposed that the incident light in the calculation has a sine-like electric field with a maximum field of 27 MV/cm and a polarization in the zigzag direction (see the Supplementary Note 5 for the case of the polarization in the armchair direction). Moreover, we assumed that the electron-hole pairs are created at the $K$ and $K'$ points when the electric field reaches a peak (at $t = 0, T/2, T, ...$ in Fig. 3d) through the Zener tunneling process. The electron-hole pairs are accelerated by the light field and eventually recombine with each other by emitting high-energy photons. When the electron-hole pairs in the $K$ valley are driven in the $K - \Gamma - K'$ direction, those in the $K'$ valley are driven in the $K' - M - K$ direction (Fig. 3b).

Here, we propose an extension of the conventional three-step model that includes quantum mechanical effects for electron-hole pairs in solids: the trajectories of electrons and holes in momentum space are calculated using the equation of motion for Bloch electrons: $\hbar \dot{\mathbf{k}} = q\mathbf{E}$, where $\mathbf{k}$ is the electron's wave vector, $q$ its charge, and $\mathbf{E}$ the electric field. In real space, we should consider a wave packet of Bloch electrons or holes. The equation of motion should be $\dot{\mathbf{r}} = \hbar^{-1} \partial \varepsilon(\mathbf{k})/\partial \mathbf{k}$. It is obvious that the wave packet can occupy any lattice site in real space due to the periodicity. Therefore, we considered several trajectories starting from different atomic sites. Figures 4a (4b) shows the real-space trajectories of the carriers generated at $t = 0$ at the $K$ ($K'$) point. The green curves indicate the motion of a hole generated at $x/a = 0$. The black dashed curves show the motion of electrons generated at several neighboring atomic sites around the hole generation site ($x/a = 0$). HHG emissions can happen when an electron and hole meet again in real space. For a hole generated at the $K$ point, there are four ways for it to recombine with the electrons, marked as circles in Fig. 4a. In this case, it should be noted that the hole recombines with an electron generated at a different atomic site. This would be a characteristic feature of HHG in solids. For the hole generated at the $K'$ point, there are also four recombination possibilities, marked as circles in Fig. 4b. By comparing the dynamics in real space with those in momentum space, one can determine the momentum and energy of electron-hole pairs at the recombination time. The circles in Fig. 4c represent the recombined electron-hole pairs generated at $t = 0$ in momentum space. The blue solid (red open) circles indicate the electron-hole pairs generated at the $K$ ($K'$) point, which result in an $\sigma^+(\sigma^-)$ emission. The two pairs generated at the $K$ points labeled 2 and 3 are in the band nesting region at the time of recombination. The band nesting causes a large joint density of states in the C and D energy region, as shown in the absorption spectra (Figs. 2a-2d). The large joint density of states in turn causes a resonant enhancement of HHG at those energies (Figs. 2e-2h). Actual electrons and holes are wave packets with a finite uncertainty in momentum, originating from the uncertainty of the time at which the pair was created via the Zener tunneling process. Thus, if we think about the electron-hole pairs in the band nesting region (red open circles labeled 2 and 3) and the others, the former cause light emissions with a narrower energy distribution than those of the latter, which cause the resonant enhancement in the band nesting energy. Figure 4d shows the same kind of plot as Fig. 4c for the electron-hole pairs generated at $t = T/2$. A half period later, the dynamics in momentum space reverse and the electron-hole pairs generated at the $K'$ point reach the band nesting region. From the viewpoint of the polarization selection rule, the alternating nature of the dynamics in the $K$ and $K'$ valleys at $t = 0$ and $t = T/2$ is important. The interband emission process changes every half period of the incident light; that is, light with the same photon energy but opposite helicity is emitted in the subsequent half period. As discussed in the previous section, the asymmetric nature of the dynamics in the $K$ and $K'$ valleys directly lead to polarization selection rules. These insights are also confirmed using a microscopic HHG model (see the Supplementary Note 4).

## Discussion

In summary, we investigated generation of high harmonics in monolayer TMDs. By comparing four monolayer TMDs, we found resonant enhancement of HHG with the interband optical transition due to band nesting effects. We also found that odd and even orders of HHG have parallel and perpendicular polarizations under excitation with polarization along with the zigzag direction. This comes from the asymmetric nature of the dynamics in the $K$ and $K'$ valleys. Our findings give the important indication that the nonlinear interband polarization significantly contributes to the high harmonic generation in solids and opens the way for attosecond science with monolayer materials having widely tunable electronic structures.

## Methods

**Samples.** The monolayer TMDs were grown on sapphire substrates by chemical vapor deposition. The monolayer flake size is typically tens of micrometers. The $WS_2$ and $WSe_2$ monolayers were prepared using the method reported in Ref. [37], and $MoS_2$ and $MoSe_2$ were purchased from 2d Semiconductors, Inc. The monolayer nature of the TMD samples was confirmed by photoluminescence and Raman spectroscopy.

**Experiments.** Supplementary Figure S1 shows the experimental configuration. A mid-infrared femtosecond laser pulse was generated using a differential frequency generation (DFG) configuration consisting of a combination of a Ti:sapphire based regenerative amplifier (800 nm center wavelength, 35 fs pulse duration, 1 kHz repetition rate, and 1 mJ pulse energy) and a $\beta$-BaB$_2$O$_4$ (BBO) based optical parametric amplifier (OPA). The photon energies of the signal and idler generated by the OPA can be tuned from 0.84 to 0.98 and from 0.71 to 0.57 eV, respectively. The difference frequency of the signal and idler beams was produced in a AgGaS$_2$ crystal, which yielded mid-infrared pulses centered at 0.12-0.41 eV. A long-pass filter transparent on the longer wavelength side above 4 μm (0.31 eV) was used to block the signal and idler beams. For the measurement of the power dependence, the power of the mid-infrared excitation pulses was controlled by using a liquid crystal variable retarder (LCC1113-MIR, Thorlabs) and a ZnSe wire grid polarizer (WP25H-Z, Thorlabs). The available photon energy of the mid-infrared pulses was restricted by the transmission photon energy of the long-pass filter and the variable retarder; it was 0.23-0.29 eV. The mid-infrared excitation pulse was focused on the sample by a ZnSe lens with a 62.5 mm focal length to a spot about 30 μm in radius. The spectral linewidth of the excitation pulse was ~60 meV in full-width half-maximum (Supplementary Figure S2**a**). Assuming a Fourier-transform-limited pulse, the pulse duration was estimated to be 25 fs. The maximum intensity of the pump pulse at the sample was 1.7 TW/cm$^2$, corresponding to an electric field of 27 MV/cm inside the monolayer TMDs, when taking into account the reflectivity of the substrate. The generated HHG was collected by a fused-silica lens (f = 50 mm) and analyzed by a grating spectrometer (iHR320, Horiba) equipped with a Peltier-cooled charge-coupled device camera (Syncerity CCD, Horiba). The high-harmonic spectra between 1.1 to 4.2 eV including the fifth to fifteenth harmonics were corrected for the grating efficiency and the quantum efficiency of the detector. The data around 1.55 eV in the spectra are not shown because there is a stray light in the experimental setup caused by the output of Ti:sapphire regenerative amplifier. The full spectrum from the monolayer $MoS_2$ up to 5 eV without an efficiency correction over 4.2 eV showed harmonics up to eighteenth order (Supplementary Figure S2**b** ). All experiments were performed at room temperature. The optical absorption spectra were measured with a scanning spectrophotometer (UV-3100PC, Shimadzu).

## Acknowledgements

This work was supported by a Grant-in-Aid for Scientific Research (A) (Grant No. 26247052) and Scienctific Research (S) (Grant No. 17H06124). N. Y. was supported by a JSPS fellowship (Grant No. 16J10537). Y.M. acknowledges support from JST CREST Grant Number JPMJCR16F3 and a Grant-in-Aid for Scientific Research(B) (no. JP18H01832) from the Ministry of Education, Culture, Sports, Science and Technology (MEXT), Japan.

## Author contributions

N. Y. and K. T. conceived the experiments. N. Y. and K. U. carried out the experiments and analyses. Y. T. S. S. and Y. M. fabricated the samples. K. N. and K. T. performed the calculation. N. Y. and K. T. wrote the manuscript. All the authors contributed to the discussion and interpretation of the results.

**Figure Captions**

**Figure 1 High harmonic generation from monolayer TMDs.** High harmonic spectra from **a** MoSe$_2$, **b** WSe$_2$, **c** MoS$_2$, and **d** WS$_2$ monolayers at room temperature induced by mid-infrared pulse excitation (photon energy 0.26 eV, peak intensity 1.7 TW/cm$^2$). We set the excitation polarization in parallel with the zigzag direction. The harmonics from fifth to sixteenth order are labeled.

**Figure 2 Resonant enhancement of even-order high harmonics.** Intensities of even-order harmonics in **a** MoSe$_2$, **b** WSe$_2$, **c** MoS$_2$, and **d** WS$_2$ monolayers. **e, f, g, h** Corresponding optical absorption spectra of the TMD monolayers. The peaks of the A and B excitons are labeled A and B, respectively. The peaks due to the band nesting effects are labeled C and D. **i, j, k, l** Corresponding intensities of odd-order harmonics.

**Figure 3 Electron-hole dynamics of HHG in monolayer TMDs. a** Top view of the crystal of monolayer TMDs. The brown dots are transition-metal atoms, while the yellow dots are chalcogenide atoms. The $x$ ($y$) axis corresponds to the zigzag (armchair) direction. **b** Two-dimensional hexagonal Brillouin zone of monolayer TMDs. The high symmetry, $\Gamma$, $K$ ($K'$), $M$ points are labeled. The electric field with a polarization in the $x$ ($y$) direction in real space drives carriers in the $k_x$ ($k_y$) direction. **c** Schematic diagrams of the real space trajectories of electrons (blue curve) and holes (red curve) driven by light field. The electron-hole pairs generated by Zener tunneling at $t = 0$ are accelerated by the electric field, and they recombine when they meet again ($t = t_R$). **d** Schematic drawings of electron-hole creation and recombination dynamics under excitation with a polarization along the zigzag direction. The electron-hole pairs generated at $t = 0$ follow different traces depending on whether they were created at the $K$ or $K'$ point. Thus, the electron-hole pairs recombine at different times $t_1$ and $t_2$ (blue solid and red open circles in the dashed orange oval). The electron-hole pairs generated at $t = T/2$ follow reversed dynamics to those created at $t = 0$ (red open and blue solid circles in the dashed green oval). **e** On the other hand, under excitation with the polarization along the armchair direction, the acceleration and recombination dynamics of the electron-hole pairs do not depend on whether the pairs were created at the $K$ point or at the $K'$ point.

**Figure 4 Semi-classical calculation of electron-hole dynamics in the real and momentum space. a** Real space dynamics of carriers generated at $t = 0$ and the $K$ point under excitation with polarization along zigzag direction. The green curve shows the trajectory of the hole generated at $x/a = 0$, and the black dashed curves show the electrons generated at $x/a = 0, -1, -2, -3, -4$. The possible recombination paths of the hole and electrons are marked as blue solid circles. **b** Same as **a** but for the carriers generated at the $K'$ point. The possible recombination paths of the hole and electrons are marked as red open circles. **c** Recombined electron-hole pairs in momentum space generated at $t = 0$. The electron-hole pairs created at the $K$ ($K'$) point are represented as blue solid (red open) circles. The labels from 1 to 4 correspond to those in **a** and **b**. The asymmetric features of the electron-hole dynamics in the $K$ and $K'$ valleys lead to different times and energies of recombination. Only the electron-hole pairs generated at the $K$ point reach the band nesting region. **d** Same as **c** but for electron-hole pairs generated at $t = T/2$. The dynamics in the $K$ and $K'$ valleys are reversed from those of $t = 0$ and show the alternating nature. Only electron-hole pairs generated at the $K$ point reach the band nesting region.

**Fig. 1**

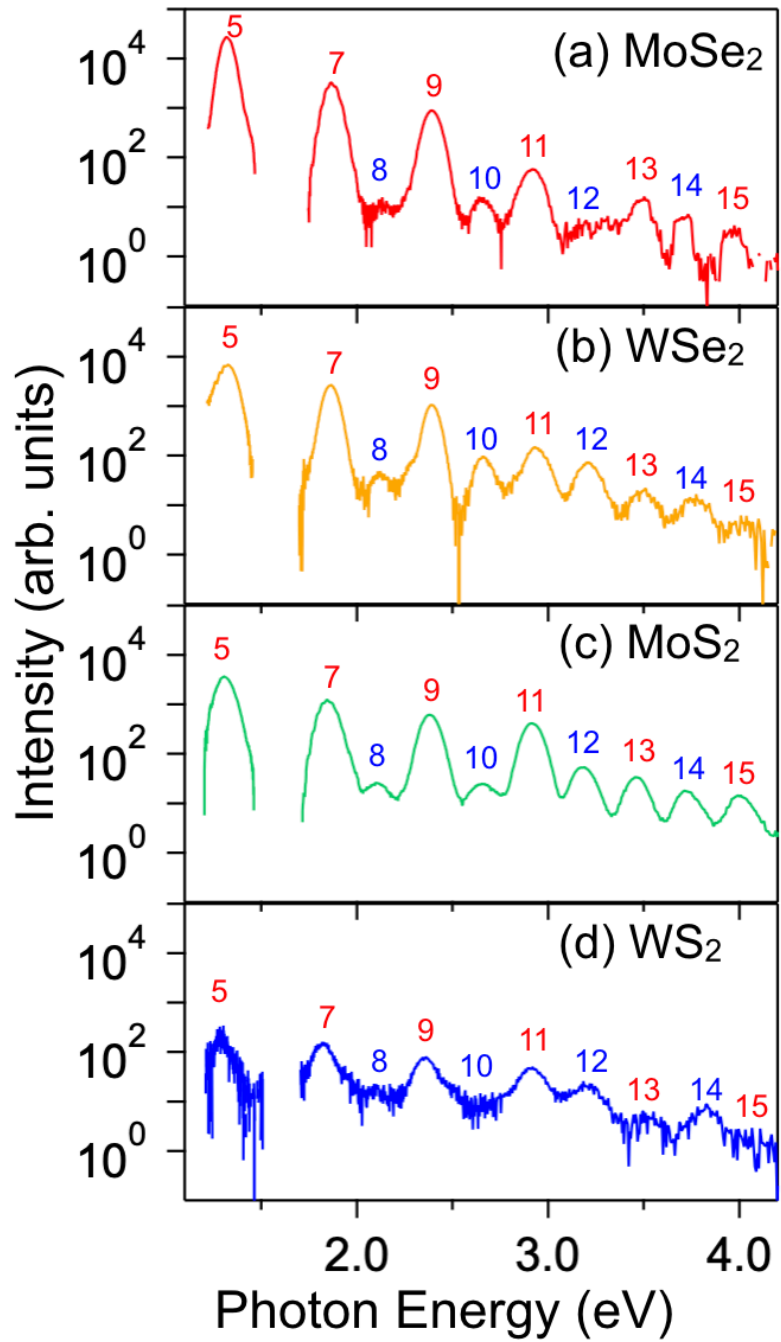

**Fig. 2**

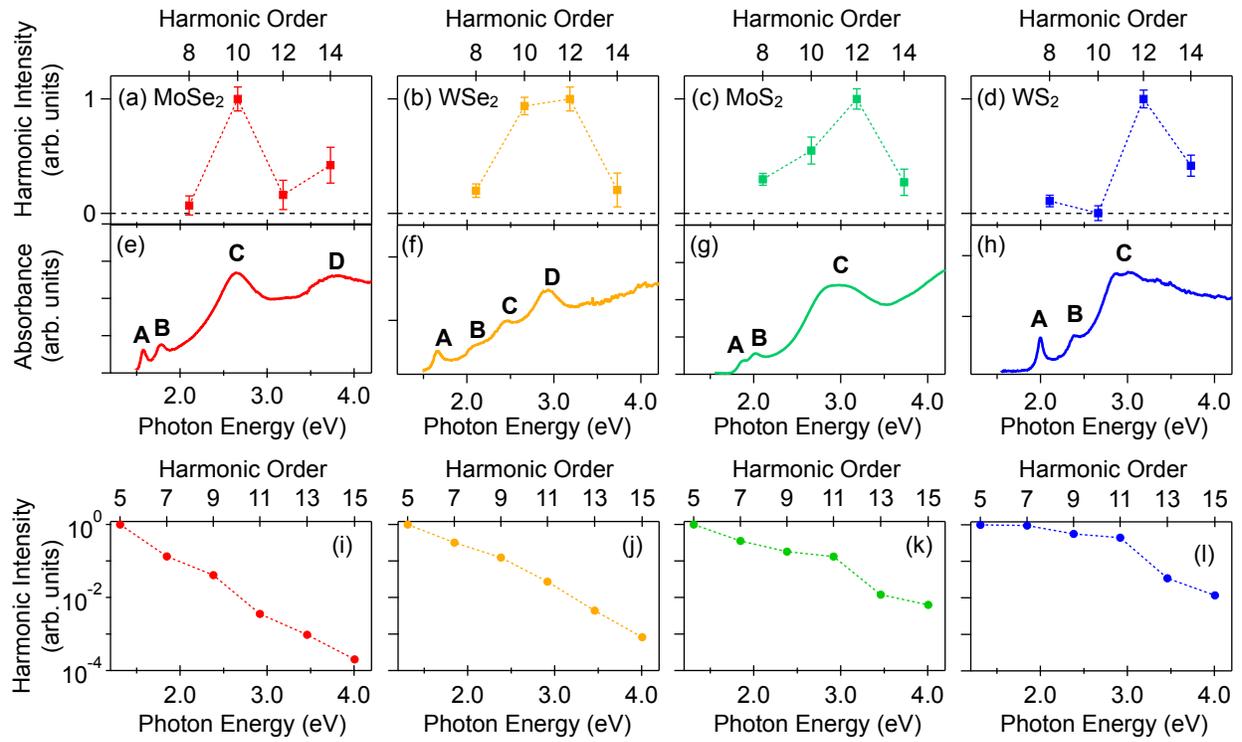

**Fig. 3**

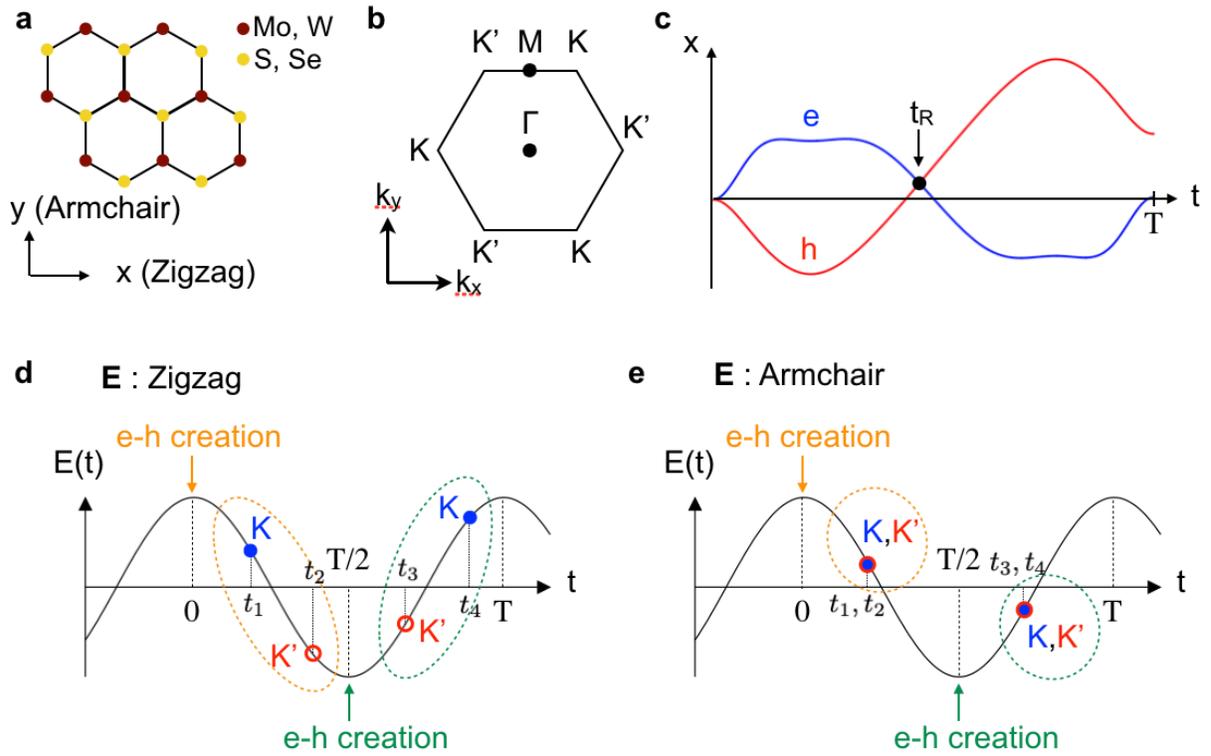

**Fig. 4**

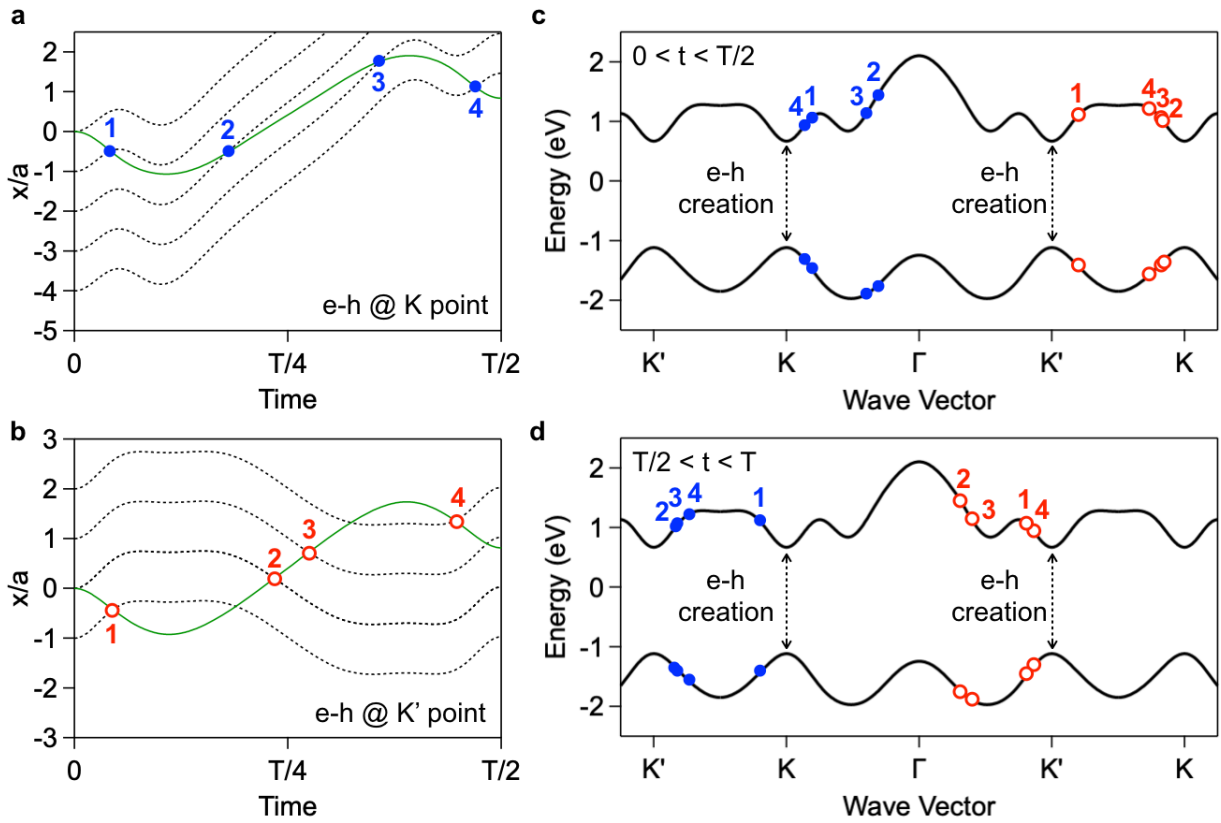

# Interband resonant high-harmonic generation by valley polarized electron-hole pairs

# —Supplementary Information—

Naotaka Yoshikawa[1,2,*], Kohei Nagai[1], Kento Uchida[1], Yuhei Takaguchi[3], Shogo Sasaki[3], Yasumitsu Miyata[3], and Koichiro Tanaka[1,2,†]

[1] *Department of Physics, Graduate School of Science, Kyoto University, Sakyo-ku, Kyoto 606-8502, Japan*

[2] *Institute for Integrated Cell-Material Sciences (WPI-iCeMS), Kyoto University, Sakyo-ku, Kyoto 606-8501, Japan*

[3] *Department of Physics, Tokyo Metropolitan University, Hachioji, 192-0397, Japan.*

[*]yoshikawa@thz.phys.s.u-tokyo.ac.jp

[†]kochan@scphys.kyoto-u.ac.jp

In this supplementary information, we provide supplementary figures for the Method section and several notes for detailed experimental results of polarization resolved HHG, power dependence, detailed description of the tight-binding band structure model, polarization selection rules with microscopic model and calculated dynamics of electron-hole pairs with the armchair excitation.

**I. Supplementary figures for the Methods section**

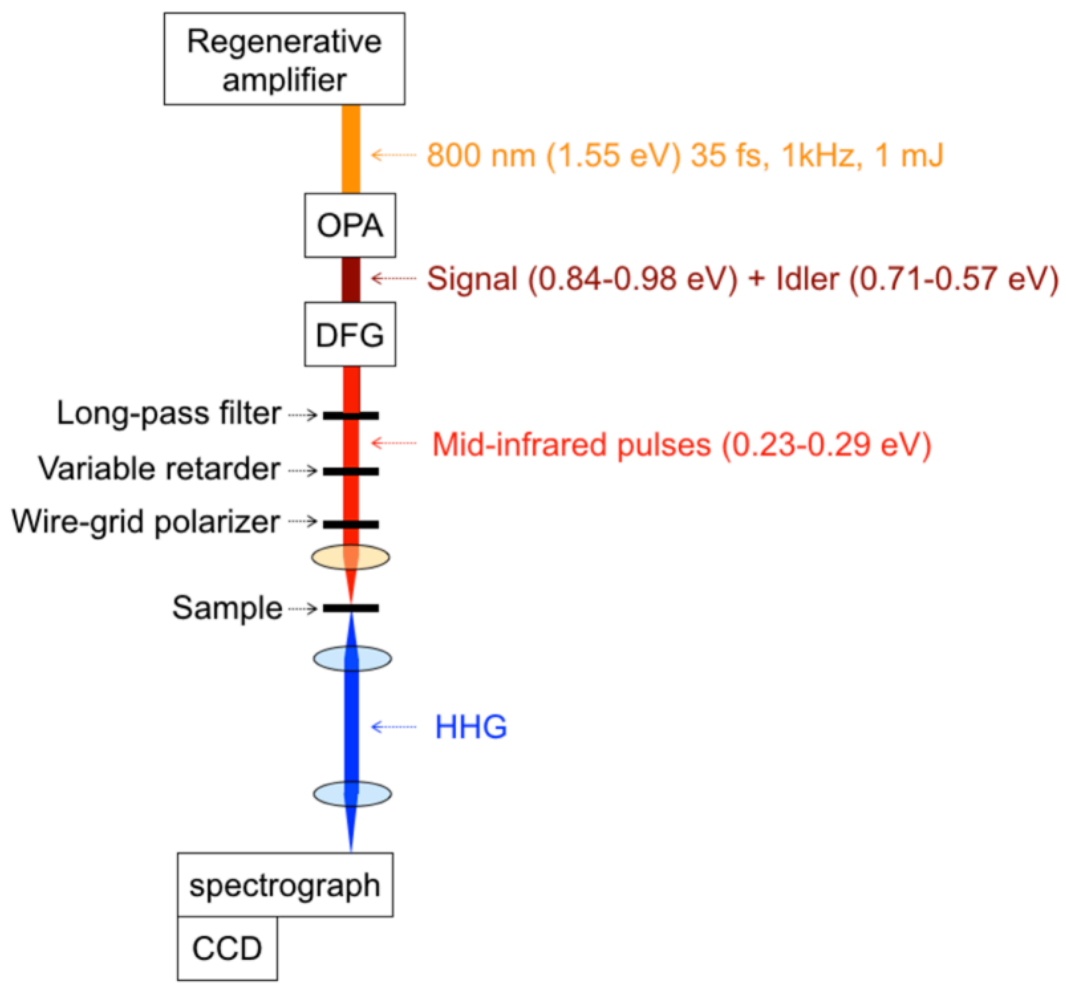

**Supplementary Figure S1** Schematic diagram of the experimental setup.

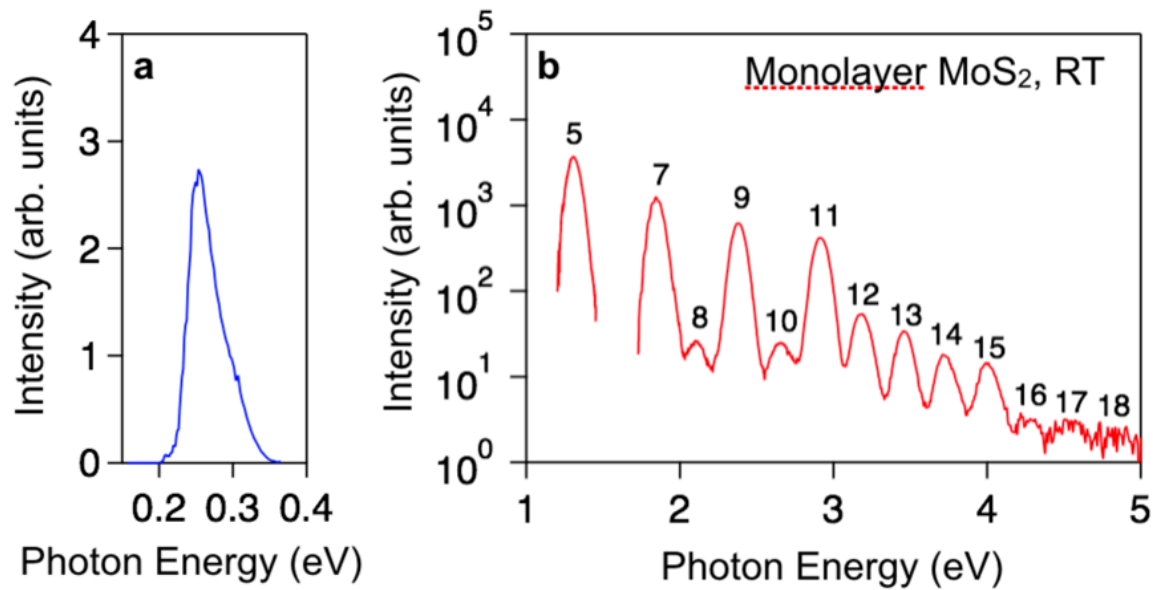

**Supplementary Figure S2** Spectrum of **a** mid-infrared excitation light and **b** HHG emitted from monolayer $MoS_2$.

## II. Supplementary Note 1: Polarization of high-harmonic generation (HHG) from monolayer MoS$_2$

The CVD samples had many crystal domains of monolayers, and their crystal orientations of the monolayer islands are not identical. We measured the harmonic spectra at the position on the sample where strong even-order high harmonics were observed; that is, the crystal orientation with respect to the polarization of the incident laser was optimized to maximize the even-order harmonics. Supplementary Figure S3 shows the polarization of the ninth and twelfth harmonics with the zigzag and armchair excitation. The odd-order harmonics are always parallel to the incident light polarization. The even-order harmonics with the zigzag excitation, where the even-order harmonics are generated efficiently, are perpendicularly polarized to the excitation light, while those with the armchair excitation are parallel. These polarization selection rules are originated from the dynamical symmetry of TMDs under the strong driving field and are discussed using the microscopic HHG model in solids in the Supplementary Note 4.

**a** Zigzag

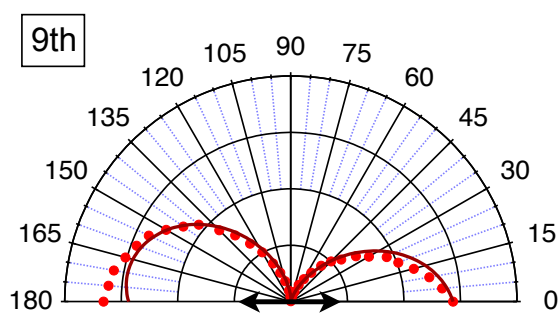 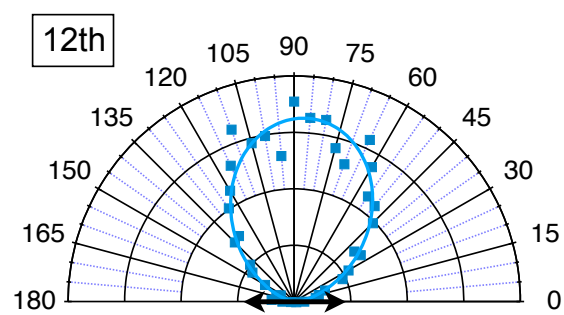

**b** Armchair

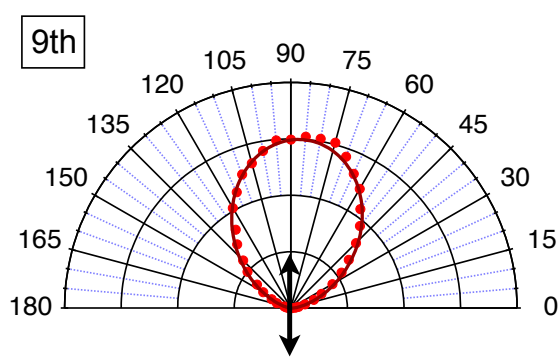 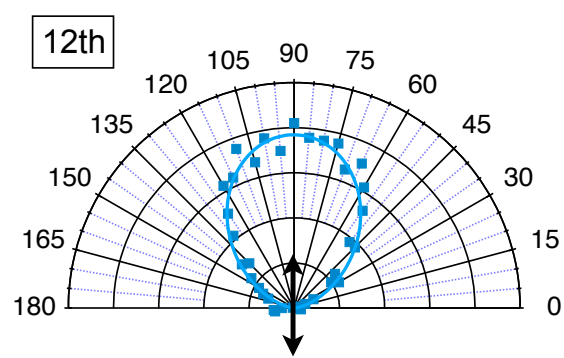

**Supplementary Figure S3** Polarization of ninth and twelfth harmonics under **a** zigzag and **b** armchair excitations in monolayer $MoS_2$. The black arrows indicate the polarization of the incident mid-infrared light.

# III. Supplementary Note 2: Nonperturbative behavior of HHG in monolayer MoS$_2$

Supplementary Figure S4 shows the intensity of the high harmonics generated from monolayer MoS$_2$ as a function of the peak power of the excitation pulses $I_{\text{exc.}}$. The power dependence shows a saturation-like behavior, whereas it should show an $I^n$ dependence ($n$ means harmonic order) in the perturbative limit. All of the observed, both even and odd, harmonics show an $\sim I^3$ dependence at the highest excitation power used in this study. The power dependence confirms the non-perturbative behavior of HHG in monolayer TMDs.

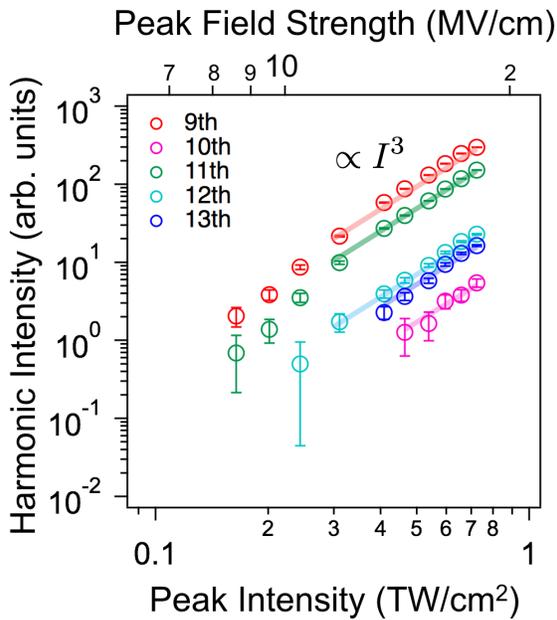

**Supplementary Figure S4 Excitation power dependence of the high harmonic intensity from monolayer MoS$_2$.** Intensities of harmonic radiation of various orders from monolayer MoS$_2$ as a function of peak intensity of the excitation. The solid lines are guides for the eye, indicating the harmonic intensities are proportional to $I^3$.

# IV. Supplementary Note 3: Band structure of monolayer MoS$_2$ determined using tight-binding model

The band structure of the monolayer MoS$_2$ used for the semi-classical calculation shown in Fig. 4 (Main Text) and Supplementary Figures S7(b) and S7(c) is given by the tight binding model developed in Suppl. Refs. [1] and [2]. The valence and conduction bands have the following equations.

$$E_\pm(\mathbf{K}) = \frac{\varepsilon_m(\mathbf{K}) + \varepsilon_s(\mathbf{K})}{2} \pm \frac{1}{2}\sqrt{[\varepsilon_m(\mathbf{K}) - \varepsilon_s(\mathbf{K})]^2 + 4|t_{sm}(\mathbf{K})|^2}, \quad \text{(S1)}$$

where

$$\varepsilon_\ell(\mathbf{K}) = \varepsilon_\ell + \sum_{\boldsymbol{\delta}>0} 2t_{\ell\ell}(\boldsymbol{\delta})\cos(\mathbf{K}\cdot\boldsymbol{\delta}), \quad \ell = s, m, \quad \text{(S2)}$$

$$|t_{sm}(\mathbf{K})| = \sqrt{3 + 2\cos K_x a + 4\cos\frac{K_x a}{2}\cos\frac{\sqrt{3}K_y a}{2}}. \quad \text{(S3)}$$

The parameters given in Suppl. Ref. [1] are summarized in **Supplementary Table 1**.

**Supplementary Table 1**. The parameters used for the band calculation taken from Suppl. Ref. [1].

| $\boldsymbol{\delta}$ | $a$ | $\sqrt{3}a$ | $2a$ | $\sqrt{7}a$ | $3a$ | $2\sqrt{3}a$ | $\sqrt{13}a$ | $4a$ |
|---|---|---|---|---|---|---|---|---|
| $t_{ss}$/meV | 45 | 15 | 60 | 0 | 5 | $-35$ | 5 | 15 |
| $t_{mm}$/meV | 20 | 100 | 10 | $-5$ | $-5$ | 10 | 0 | $-5$ |

$\varepsilon_m = 1.15$ eV, $\varepsilon_s = -1.7$ eV, $t_{sm} = 0.3$ eV

By using the obtained tight-binding band structure, we also calculated the group velocity $\dot{\mathbf{r}} = \hbar^{-1}\partial\varepsilon(\mathbf{k})/\partial\mathbf{k}$, where $\mathbf{r}$ is the carrier's position and $\mathbf{k}$ its wave vector, as shown in Supplementary Figure S5. This result was used in the calculation of the motion of the electrons and holes in Fig. 4 in the main text. Supplementary Figure S6 shows the calculated joint density of states along the $K - \Gamma - K'$ trajectory in the reciprocal space. It shows divergence features at 1.8, 2.8, 2.9, and 3.3 eV. The peak at 1.8 eV is derived from the bandgap at $K$ and $K'$ points, and that at 3.3 eV is derived from the $\Gamma$ point. The peaks at 2.8 and 2.9 eV corresponds to the band nesting region where the optical absorption and high harmonic intensity have the resonant enhancement (Fig. 2 in the main text).

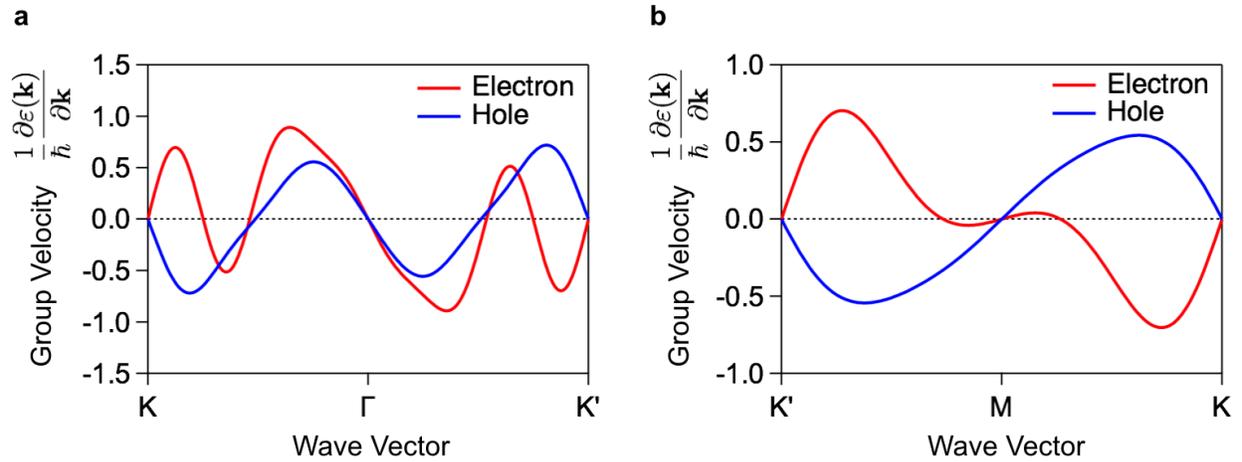

**Supplementary Figure S5** Group velocity of electrons and holes in monolayer MoS$_2$ for **a** $K - \Gamma - K'$ and **b** $K - \mathrm{M} - K'$.

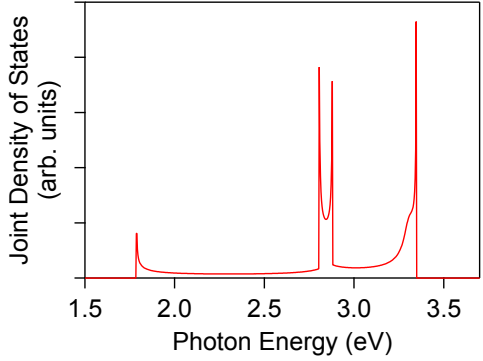

**Supplementary Figure S6** Joint density of states of monolayer MoS$_2$ calculated along the $K - \Gamma - K'$ trajectory in the reciprocal space.

# V. Supplementary Note 4: Symmetry analysis of polarization selection rules

In the main text, the polarization selection rules of HHG with the zigzag excitation are discussed. Here, we derive the polarization selection rule from the microscopic model which describes the three step HHG mechanism in solids. Using the formulation discussed in Suppl. Ref. [3], the interband polarization density $\boldsymbol{p}(t)$ can be written in the length gauge as follows:

$$\boldsymbol{p}(\text{t}) = \frac{i}{\hbar} \int_{BZ} d^2 \boldsymbol{k}\, \boldsymbol{\mu}^*(\boldsymbol{k}) \frac{i}{\hbar} \int_{-\infty}^{t} dt'\, \boldsymbol{\mu}\big(\boldsymbol{k}(t,t')\big) \cdot \boldsymbol{E}_{MID}(t') \exp\left(-i \int_{t'}^{t} dt''\, \varepsilon_g\big(\boldsymbol{k}(t,t'')\big)/\hbar\right) + \text{c.c.},$$

(S4)

$$\boldsymbol{k}(t,t') = \boldsymbol{k} + \frac{e}{\hbar} \boldsymbol{A}(t') - \frac{e}{\hbar} \boldsymbol{A}(t).$$

(S5)

Here, $\boldsymbol{k}$ is the crystal momentum, $\boldsymbol{\mu}$ is the transition dipole moment, $\varepsilon_g$ is the transition energy from the valence to the conduction band, and $\boldsymbol{A}(t)$ is the vector potential of the driving field satisfying $\boldsymbol{E}_{MID}(t) = -\partial \boldsymbol{A}(t)/\partial t$. The equation (S4) describes the induced dipole moment for HHG including electron-hole pair creation through the tunneling process and electron-hole motion due to the driving field. For simplicity, we do not account for the spin-orbit coupling in this model. In order to derive the selection rule for HHG, we divide $\boldsymbol{p}(t)$ into two terms given by

$$\boldsymbol{p}(t) = p_+(t)\boldsymbol{\sigma}^+ + p_-(t)\boldsymbol{\sigma}^-, \tag{S6}$$

$$p_\pm(t) = \frac{i}{\hbar}\int_{BZ} d^2k\, \mu_\pm^*(\boldsymbol{k}) \int_{-\infty}^t dt'\, \boldsymbol{\mu}\bigl(\boldsymbol{k}(t,t')\bigr)\cdot \boldsymbol{E}_{MID}(t')\exp\left(-i\int_{t'}^t dt''\, \varepsilon_g\bigl(\boldsymbol{k}(t,t'')\bigr)/\hbar\right) + \text{c. c.}, \tag{S7}$$

Then, let us consider continuous-wave electric field along zigzag direction, where $\boldsymbol{E}_{zigzag}^{MID}(t) = E_x\cos(\Omega t)\,(\boldsymbol{\sigma}^+ + \boldsymbol{\sigma}^-)/\sqrt{2}$, and the temporal-translation of the interband polarization $\boldsymbol{p}(t+T/2)$ ($T = 2\pi/\Omega$) is given by

$$p_\pm\left(t+\frac{T}{2}\right)$$
$$= -\frac{i}{\hbar}\int_{BZ} d^2k\, \mu_\pm^*(-k_x, k_y) \int_{-\infty}^t dt'\, \mu_x(-k_x(t,t'), k_y) E_x\cos(\Omega t')\exp\left(-i\int_{t'}^t dt''\, \varepsilon_g(-k_x(t,t''), k_y)/\hbar\right)$$
$$+\text{c. c.}, \tag{S8}$$

$$k_i(t,t') = k_i + \frac{e}{\hbar}A_i(t') - \frac{e}{\hbar}A_i(t). \quad (i = x \text{ and } y) \tag{S9}$$

$$\mu_x = (\mu_+ + \mu_-)/\sqrt{2}. \tag{S10}$$

Since the electronic system in TMDs without the driving field $\boldsymbol{E}_{zigzag}^{MID}(t)$ has a mirror symmetry plane normal to the x-axis (zigzag axis), the transition energy $\varepsilon_g$ and the dipole moment $\boldsymbol{\mu}$ satisfy

$$\varepsilon_g(-k_x, k_y) = \varepsilon_g(k_x, k_y), \tag{S11}$$

$$\mu_{\pm}(-k_x, k_y) = -\mu_{\mp}(k_x, k_y). \tag{S12}$$

These conditions do not supply any relation between $p_+(t)$ and $p_-(t)$ but give a relation between $\boldsymbol{p}(t + T/2)$ and $\boldsymbol{p}(t)$ as follows:

$$p_{\pm}\left(t + \frac{T}{2}\right)$$

$$= -\frac{i}{\hbar} \int_{BZ} d^2k\, \mu_{\mp}^*(k_x, k_y) \int_{-\infty}^{t} dt'\, \mu_x(k_x(t,t'), k_y) E_x \cos(\Omega t') \exp\left(-i \int_{t'}^{t} dt''\, \varepsilon_g(k_x(t,t''), k_y)/\hbar\right)$$

$$+ \text{c. c.}$$

$$= -p_{\mp}(t). \tag{S13}$$

Since electric field of high-harmonics is approximately given by $\boldsymbol{E}^{HHG}(t) \propto d^2\boldsymbol{p}(t)/dt^2$, we can obtain Eq. (2) in the main text ($E_{\pm}^{\text{HHG}}(t) = -E_{\mp}^{\text{HHG}}(t + \frac{T}{2})$), resulting in parallel (x-polarized) odd-order harmonics and perpendicular (y-polarized) even-harmonics that we observed in the experiments.

Next, we consider the polarization of HHG with the armchair excitation. The electric field of the excitation is described with the polarization of the armchair direction (y axis) of the mid-infrared excitation, as $\boldsymbol{E}_{armchair}^{MID}(t) = E_y \cos(\Omega t)\,(\boldsymbol{\sigma}^+ + \boldsymbol{\sigma}^-)/\sqrt{2}i$. In contrast to the zigzag excitation case, one can find a simple relation between $p_+(t)$ and $p_-(t)$ since the driving field is within the mirror symmetry plain. Using the equations (S11) and (S12), the relation between $p_+(t)$ and $p_-(t)$ is given by

$p_\pm(t)$

$$= \frac{i}{\hbar} \int_{BZ} d^2k\, \mu_\pm^*(-k_x, k_y) \int_{-\infty}^{t} dt'\, \mu_y\left(-k_x, k_y(t,t')\right) E_y \cos(\Omega t') \exp\left(-i \int_{t'}^{t} dt''\, \varepsilon_g\left(-k_x, k_y(t,t'')\right)/\hbar\right)$$

$+$ c. c.

$$= -\frac{i}{\hbar} \int_{BZ} d^2k\, \mu_\mp^*(k_x, k_y) \int_{-\infty}^{t} dt'\, \mu_y\left(k_x, k_y(t,t')\right) E_y \cos(\Omega t') \exp\left(-i \int_{t'}^{t} dt''\, \varepsilon_g\left(k_x, k_y(t,t'')\right)/\hbar\right)$$

$+$c. c.

$$= -p_\mp(t), \tag{S14}$$

$$\mu_y = (\mu_+ - \mu_-)/\sqrt{2}i. \tag{S15}$$

As a result, one can deduce the rules:

$$E_\pm^{HHG}(t) = -E_\mp^{HHG}(t) \tag{S16}$$

Equation (S16) leads to the following equation:

$$E_x^{HHG}(t) = 0, \tag{S17}$$

This means that both even- and odd-order HHG emission appear with the parallel polarization (y-polarized) in the case of armchair excitation.

We briefly touch the reason why we can obtain the parallel-polarized even-harmonics although the motions of electron-hole pairs in the ranges $0 < t < T/2$ and $T/2 < t < T$ are symmetric in the case of armchair excitation. The temporal-translation of the interband polarization $p(t + T/2)$ is given by

$$p_\pm\left(t+\frac{T}{2}\right) =$$

$$-\frac{i}{\hbar}\int_{BZ} d^2k\, \mu_\pm^*(k_x,-k_y) \int_{-\infty}^{t} dt'\, \mu_y\left(k_x,-k_y(t,t')\right) E_y\cos(\Omega t') \exp\left(-i\int_{t'}^{t} dt''\, \varepsilon_g\left(k_x,-k_y(t,t'')\right)/\hbar\right) +$$

c. c. (S18)

Since electronic system in TMDs has the time reversal symmetry, transition energy $\varepsilon_g$ and dipole moment $\mu$ satisfy

$$\varepsilon_g(-\boldsymbol{k}) = \varepsilon_g(\boldsymbol{k}), \tag{S19}$$

$$\mu_\pm(-\boldsymbol{k}) = \mu_\mp^*(\boldsymbol{k}). \tag{S20}$$

By combining equations (S11) and (S12) from the mirror symmetry, one can obtain

$$\varepsilon_g(k_x,-k_y) = \varepsilon_g(k_x,k_y), \tag{S21}$$

$$\mu_\pm(k_x,-k_y) = -\mu_\pm^*(k_x,k_y). \tag{S22}$$

By using these relations, $\boldsymbol{p}(t+T/2)$ becomes

$$p_\pm\left(t+\frac{T}{2}\right)$$

$$= \frac{i}{\hbar}\int_{BZ} d^2k\, \mu_\pm(k_x,k_y) \int_{-\infty}^{t} dt'\, \mu_y^*\left(k_x,k_y(t,t')\right) E_y\cos(\Omega t') \exp\left(-i\int_{t'}^{t} dt''\, \varepsilon_g\left(k_x,k_y(t,t'')\right)/\hbar\right)$$

+ c. c.

$$= \frac{i}{\hbar}\int_{BZ} d^2k\, |\mu_\pm(\boldsymbol{k})| \int_{-\infty}^{t} dt'\, |\mu_y(\boldsymbol{k}(t,t'))| e^{i\{\theta_\pm(\boldsymbol{k})-\theta_y(\boldsymbol{k}(t,t'))\}} E_y\cos(\Omega t') \exp\left(-i\int_{t'}^{t} dt''\, \varepsilon_g(\boldsymbol{k}(t,t''))/\hbar\right)$$

+c. c., (S23)

$$\mu_i = |\mu_i|e^{i\theta_i}. \tag{S24}$$

On the other hand, $\boldsymbol{p}(t)$ can be written by

$$p_\pm(t) =$$

$$\frac{i}{\hbar}\int_{BZ} d^2\boldsymbol{k}\, |\mu_\pm(\boldsymbol{k})| \int_{-\infty}^{t} dt'\, |\mu_y(\boldsymbol{k}(t,t'))| e^{-i\{\theta_\pm(\boldsymbol{k})-\theta_y(\boldsymbol{k}(t,t'))\}} E_y \cos(\Omega t') \exp\left(-i\int_{t'}^{t} dt''\, \varepsilon_g(\boldsymbol{k}(t,t''))/\hbar\right) +$$

$$c.c.. \tag{S25}$$

From the equations (S23) and (S24), it is clear that $\boldsymbol{p}(t+T/2)$ and $\boldsymbol{p}(t)$ have a difference in the phase of the complex dipole moment such as $\{\theta_\pm(\boldsymbol{k}) - \theta_y(\boldsymbol{k}(t,t'))\}$, permitting the even-orders of HHG in the parallel polarization with the driving field, even though motions of the electron-hole pair are symmetric in the case of armchair excitation.

## VI. Supplementary Note 5: Calculated dynamics of electrons and holes under mid-infrared excitation with the armchair polarization

In the main text, we used the extended three-step model to examine the dynamics of electrons and holes with the zigzag-polarization excitation. Here, we discuss the dynamics for the armchair-polarization excitation. Supplementary Figure S7**a** shows the group velocity map of the conduction band electrons in momentum space, as calculated with the tight binding model. With the armchair excitation (in the k_y direction represented by the orange arrows), the wave packets of Bloch electrons are accelerated not only in the y direction but also in the x direction in real space, whereas they are accelerated only in the x direction with the zigzag excitation (in the k_x direction represented by the green arrows). We regard a collision of an electron and hole as a pair whose separation is smaller than 0.05 $a$. Here, we assume that these electron-hole pairs are generated at the $K$ and $K'$ points at $t = 0$, at which the positive peak of the incident light field occurs. We calculated the dynamics of a hole generated at $(x, y) = (0,0)$ and several electrons generated at 60 neighboring atomic sites in addition to the atomic site of the hole. We limited the collision time to a half cycle of the incident light field ($0 < t < T/2$). The two circles in Supplementary Figures S7**b** and S7**c** show the recombined electron-hole pairs in momentum space. Even with the armchair polarization, the electrons and holes recombine and HHG is emitted. It is clear that the collision dynamics of the electrons and holes generated at the $K$ points are equivalent to those at the $K'$ points as is clearly shown in Supplementary Figures S7**b** and S7**c**. This situation is reproduced by the theoretical model described in the Supplementary Note 4. However, we should also pay attention to the result that the phases of induced polarizations are different by $\pi$ as shown in Eqn. (S14) in Supplementary Note 4. These results clearly predict that no HHG is observed with perpendicular polarization (zigzag) in the case of the armchair excitation, which is experimentally confirmed in Supplementary Figure S3.

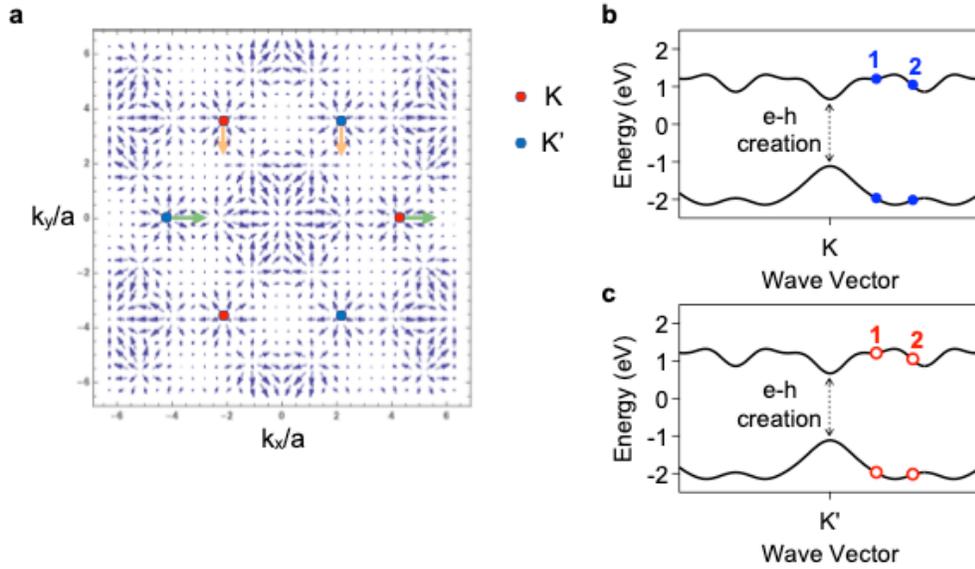

**Supplementary Figure S7 Calculation of dynamics for the armchair polarization. a** Group velocity map of conduction band electrons calculated using tight-binding model. Orange (green) arrows represent the acceleration direction in the armchair (zigzag) polarization. **b, c** Recombined electron-hole pairs generated at $K$ and $K'$ in momentum space along armchair direction. The recombined electron-hole pairs created at the $K$ ($K'$) point are represented as blue solid (red open) circles. The labels 1 and 2 indicate the two possible recombination paths at $t = 0.09T$ and $t = 0.26T$, respectively.

## Supplementary References